\documentclass[aps,prc,superscriptaddress,showpacs,floatfix]{revtex4-1}

\usepackage{epsfig}
\usepackage{graphicx}
\usepackage{amsmath}
\usepackage{multirow}
\usepackage{graphicx}
\usepackage{amsmath}
\usepackage{feynmf}
\usepackage[normalem]{ulem}  
\usepackage[dvips]{color} 

\renewcommand\sout{\bgroup \color{red} \ULdepth=-.5ex \ULset}

\newcommand{\Ex}[2]{\ifmmode{#1\times10^{#2}}\else{$#1\times10^{#2}$}\fi}

\begin{document}
\title{Symmetry energy in cold dense matter}

\author{Kie Sang Jeong}
\email[]{k.s.jeong@yonsei.ac.kr}
\author{Su Houng Lee}
\email[]{suhoung@yonsei.ac.kr}

\affiliation{Department of Physics and Institute of Physics and
Applied Physics, Yonsei University, Seoul 120-749, Korea}

\date{\today}
\begin{abstract}
We calculate the symmetry energy in cold dense matter both in the
normal quark  phase and in the 2-color superconductor (2SC) phase.
For the normal phase, the thermodynamic potential is calculated by
using hard dense loop (HDL) resummation to leading order, where the
dominant contribution comes from the longitudinal gluon rest mass.
The effect of gluonic interaction to the symmetry energy, obtained
from the thermodynamic potential, was found to be small.
 In the 2SC phase, the non-perturbative BCS paring
gives enhanced symmetry energy as the gapped states are forced to be
in the common Fermi sea reducing the number of available quarks that
can contribute to the asymmetry. We used high density effective
field theory to estimate the contribution of gluon interaction to
the symmetry energy. Among the gluon rest masses in 2SC phase, only
the Meissner mass has iso-spin dependence although the magnitude is
much smaller than the Debye mass.  As the iso-spin dependence of
gluon rest masses is even smaller than the case in the normal phase,
we expect that the contribution of gluonic interaction to the
symmetry energy in the 2SC phase will be minimal. The different
value of symmetry energy in each phase will lead to  different
prediction for the particle yields in heavy ion collision
experiment.
\end{abstract}

\maketitle

\section{Introduction}
There is a world wide interest in symmetry energy as it provides a
key to understanding the iso-spin asymmetric property of the nuclear
matter. Furthermore, related topics range from rare isotopes at low
energy regime to neutron star core at high energy regime
\cite{Steiner:2004fi, Li:2008gp}. The value of the symmetry energy
at normal nuclear matter can be inferred from either the binding
energy of semi empirical mass formula \cite{Chen:2011ek} or
experimental information from isotopes or heavy ion collision
\cite{Li:2008gp}. However, the value at higher density  is still
under debate. Some model calculations including iso-spin dependent
interaction channel (NL$\rho \delta$) \cite{Baran:2004ih} predict a
stiff symmetry energy as density becomes higher, which can cause
less favourable condition for the neutron and subsequently decreased
iso-spin density.

The isospin asymmetric matter could persist longer in a heavy ion
collision if the high density phase allows for a  quark and hadron
mixed phase. In such cases, it was found that the iso-spin density
can remain  high  even if the hadronic part of the symmetry energy
rises stiffly at high density \cite{Di Toro:2006pq, DiToro:2009ig,
Pagliara:2010ii}. This will lead to the enhanced iso-spin rich
resonances and subsequently to the increased iso-spin asymmetric
decays. This means that the particle yields can be modified not only
by the nuclear symmetry energy but also by the quark part. As
representative studies, the MIT bag model was used to calculate the
symmetry energy when the quark phase is normal \cite{Di Toro:2006pq,
DiToro:2009ig}, and additional color superconducting effect was
taken into account through an NJL type model
calculation~\cite{Pagliara:2010ii}. Also the confined iso-spin
density dependent mass (CIDDM) model was used in
Ref.~\cite{Chu:2012rd} to obtain the quark matter symmetry energy
and related influence in the quark star.

In the extremely dense condition, one should consider the equation of state
for quark matter. The equation of state for non-interacting  quarks with up to three flavors was first introduced in Ref.~\cite{Itoh:1970uw}  in relation to the hypothetical quark stars. Due to asymptotic freedom, methods based on
perturbative quantum chromodynamics (QCD) can be applied to
calculate physical observables in extreme condition. If there exists
some hard excitation scale  in the matter, soft excitations can have
infinite number of hard loop corrections that are not suppressed by
the order of coupling and that can be resumed to an equivalent order
\cite{Braaten:1989mz,Braaten:1989kk, Blaizot:1993be}. Considering
the cold normal dense matter first, in-medium quark excitation scale
is hard due to its large chemical potential, whereas the gluon
excitation scale is mainly soft. Therefore, all equivalent gluonic
excitations which contains hard dense quark loop corrections should
be resumed \cite{Manuel:1995td}. Secondly, when temperature becomes
even lower, all the fermions will be confined below their Fermi sea
without large fluctuation. As the momentum fluctuation scales to a
small value, most of the operators can be absorbed into the energy
term or neglected. On the other hand, the four-fermion interaction
with opposite momenta becomes relevant \cite{Wilson:1971bg,
Wilson:1971dh, Polchinski:1992ed, Shankar:1993pf}. If the
interaction is attractive, it is natural for the fermions to form a
condensate which leads to superconductivity: BCS paring
\cite{Bardeen:1957mv}. For QCD matter, as the quarks in the color
anti-triplet channel are attractive, color superconductivity arises
naturally with large gap size \cite{Alford:1997zt, Alford:2007xm,
Son:1998uk, Schafer:1999jg, Huang:2003xd}. At the density regime
where the quark-hadron mixed phase may exist, 2-color
superconductivity (2SC) will be favored as there is a mismatch of
Fermi momentum between light quarks and strange quark
\cite{Rajagopal:2000wf}.

In this work, we calculated the symmetry energy in cold dense matter
using QCD. The density region being considered lies between 3 to 5
times the nuclear matter density $\rho_0 = 0.16~\textrm{fm}^{-3}$,
which could be reached in heavy ion collision experiment. For the
normal phase, we use the hard dense loop (HDL) resummation for the
gluonic interaction. Thermodynamic potential is obtained to leading
order in HDL perturbation, which is an extension/modifying of
previously reported works \cite{Andersen:1999sf, Baier:1999db,
Andersen:2002jz}. The symmetry energy has been calculated from the
thermodynamic potential. For color superconductor phase, 2SC phase
is considered. Without perturbative gluonic interaction, we adopted
thermodynamic potential in 2SC phase from Ref.~\cite{Schafer:1999fe,
Miransky:1999tr} and calculated symmetry energy. To estimate the
contribution of perturbative gluonic interaction, iso-spin
dependence of gluon rest masses is calculated
 by using high density
effective theory (HDET) \cite{Hong:1998tn, Hong:1999ru,
Casalbuoni:2001ha, Nardulli:2002ma}.

This paper is organized as follows: in Sec. II, brief summary for
quark and gluon modification by HDL resummation, the thermodynamic
potential and the symmetry energy in the normal phase are presented.
In Sec. III, the gluon rest masses in iso-spin asymmetric 2SC phase
is calculated by following HDET and its contribution to the symmetry
energy is estimated. Discussion and conclusions are given in Sec.
IV.

\section{Symmetry energy at low temperature}
\subsection{Nuclear symmetry energy from equation of state}
Discussion for symmetry energy can start from a finite nuclei with
$A$ nucleons \cite{Chen:2011ek}. The Bethe-Weizs\"{a}ker formula for
the nuclear binding energy is given as
\begin{align}
m_{\textrm{tot}}&=Nm_n+Zm_p-E_B/c^2,\nonumber\\
E_B&=a_VA-a_SA^\frac{2}{3}-a_C(Z(Z-1))A^{-\frac{1}{3}}-a_AI^2A+\delta(A,Z),\label{semp}
\end{align}
where $I = (N-Z)/A$.  The fourth term in $E_B$ accounts for the
total shifted energy due to the neutron number excess. The
coefficient of fourth term $a_A$ represents the shifted energy per
nucleon.

When the atomic number $A$ becomes large, one can approach a limit where parameters describing the system become continuous
and a mean field type of approach appropriate.  In such a system, the
symmetry energy can be defined from energy per nucleon number:
\begin{align}
\frac{E(\rho_N,I)}{A}&\equiv\bar{E}(\rho_N,I)=E_0(\rho_N)+E_{\textrm{sym}}(\rho_N)I^2 + O(I^4)+\cdots,\\
E_{sym}(\rho_N)&= \frac{1}{2!} \frac{\partial^2}{\partial
I^2}\bar{E}(\rho_N,I), \label{def-sym}
\end{align}
where $A$ is the atomic number, $\rho_N$ the nuclear medium density
and $I$ the asymmetric parameter $I = (N-Z)/A \rightarrow
(\rho_n-\rho_p)/(\rho_n+\rho_p)$. The neutron and proton densities
can be found as $\rho_n = \frac{1}{2} \rho_N(1+ I)$, $\rho_p =
\frac{1}{2} \rho_N(1- I)$ respectively.

\subsection{HDL resumed thermodynamic potential}
In this section, we work in the imaginary time formalism ($t\rightarrow
-i \tau$, $0\leq\tau\leq\beta$) following Ref.~\cite{Le
Bellac:1996}. The partition function $\mathcal{Z}_\Omega$ can be
obtained in QCD degree of freedom:
\begin{align}
\mathcal{Z}_\Omega=\textrm
{Tr}\exp{\left[-\beta(\hat{H}-\vec{\mu}\cdot\vec{\hat{N}})\right]}=\int
\mathcal{D}(\bar{\psi},\psi,A,\eta)\exp\left[-\int_0^\beta d\tau
\int d^3x \mathcal{L}_E(\bar{\psi},\psi,A,\eta) \right],\label{part}
\end{align}
where the Euclidean QCD Lagrangian $\mathcal{L}_E$ is given as
\begin{align}
\mathcal{L}_E =&~
\frac{1}{4}F^a_{\mu\nu}F^a_{\mu\nu}+\frac{1}{2\xi}(\partial_\mu
A^a_\mu)^2 + \bar{\eta}^a(\partial^2
\delta_{ab}+gf_{abc}\partial_\mu
A^c_\mu)\eta^b\nonumber\\
&+\sum_f^{n_f}\left[\psi^\dagger_f\partial_\tau \psi_f+\bar{\psi}_f
(-i \gamma^i
\partial_i+m_f)\psi_f-\mu_f\psi^\dagger_f\psi_f-g\bar{\psi}_f \slash \hspace{-0.2cm}  A
\psi_f \right],\label{lagrangian}
\end{align}
where $\eta^a$ is ghost fields and the subscript $f$ represents the
quark flavor. From Eq.~\eqref{part} and Eq.~\eqref{lagrangian}, the free
propagators of fields can be obtained as
\begin{align}
S_F(P)&=\frac{m_f-\slash \hspace{-0.2cm}
P}{-(i\omega_n+\mu_f)^2+\vec{p}^2+m_f^2},\\
D^{ab}_{F,\mu\nu}(Q)&= \delta^{ab}\left[
\frac{1}{Q^2}\left(\delta_{\mu\nu}-\frac{Q_\mu
Q_\nu}{Q^2}\right)+\frac{\xi}{Q^2}\frac{Q_\mu
Q_\nu}{Q^2} \right] \nonumber\\
&= \delta^{ab}\left[ \frac{1}{Q^2}\left( P^L_{\mu\nu} +P^T_{\mu\nu}
\right) + \frac{\xi}{Q^2}\frac{Q_\mu Q_\nu}{Q^2} \right],
\end{align}
where $S_F(P)$ is the free quark propagator, $D^{ab}_{F,\mu\nu}(Q)$
the free gluon propagator and $\xi$ the gauge fixing term. The
longitudinal and transverse projection operators are defined as
\begin{align}
P^T_{ij}&=\delta_{ij}-\hat{q}_i\hat{q}_j,~~
P^T_{44}=P^T_{4i}=0,\nonumber\\
P^L_{\mu\nu}&=\delta_{\mu\nu}-\frac{Q_\mu Q_\nu}{Q^2}-P^T_{\mu\nu}.
\end{align}
In the imaginary formalism, the energy integration at each loop changes
into a sum over discrete Matusbara frequencies:
\begin{align}
\int \frac{d^4 k}{(2\pi)^4}\Rightarrow \int \frac{d^4 K}{(2\pi)^4} =
T \sum_n \int \frac{d^3 k}{(2\pi)^3},
\end{align}
where $T\sum_n$   is over $i\omega_n= i (2n) \pi T$ and
$i\tilde{\omega}_n= i (2n+1) \pi T + \mu_f$ for bosons and fermions,
respectively.

\subsubsection{HDL resumed propagator}

Before calculating the gluonic interaction to the free energy, we
summarize the medium modification of the propagator. First, we
consider  the gluon case. Suppressing light quark masses and soft
external momenta ($m_f/\mu_f \simeq 0,~ K_\mu -  Q_\mu \simeq K_\mu
$) against the hard scale of internal loop ($K_\mu \sim T, \mu_f$),
the gluon polarization tensor in the dense matter can be calculated
as \cite{Braaten:1989mz, Braaten:1989kk, Blaizot:1993be}
\begin{align}
\Pi^{ab}_{\mu\nu}(Q)&=g^2 \delta^{ab} \int \frac{d^4 K}{(2\pi)^4}
\textrm{Tr}\left[\gamma_\mu S_F(K)
\gamma_\nu S_F(K-Q)\right]\nonumber\\
& = m^2 \delta^{ab} \int
\frac{d\Omega}{4\pi}\left(\delta_{\mu4}\delta_{\nu4}+\hat{K}_\mu\hat{K}_\nu\frac{i\omega}{Q\cdot\hat{K}}\right),
\end{align}
where $\hat{K}_\mu=(-i,\hat{k}=\vec{k}/\vert \vec{k} \vert)$ is
light-like four-vector, $Q_\mu=(-\omega,\vec{q})$ is the Euclidean gluon
external momentum and $m^2$ is defined as
\begin{align}
m^2&=\frac{1}{3}g^2T^2\left(C_A+\frac{1}{2}n_f\right)+\frac{1}{2}g^2\sum_f\frac{\mu_f^2}{\pi^2}.
\end{align}
At low temperature, only HDL contribution becomes relevant: $m^2
\Rightarrow \frac{1}{2}g^2\sum_f \mu_f^2/\pi^2 \sim g^2 \mu_f^2$
\cite{Manuel:1995td}. The polarization tensor can be decomposed into
loop structures or polarizations as follows:
\begin{align}
\Pi^{ab}_{\mu\nu}(Q) & = \delta^{ab}
\left(\Pi_{\mu\nu}(Q)\vert_{\textrm{q-h}} +
\Pi_{\mu\nu}(Q)\vert_{\textrm{q-a}}\right)\\
&= \delta^{ab} \left(\delta \Pi^L(Q) P^L_{\mu\nu}+ \delta \Pi^T(Q)
P^T_{\mu\nu}\right).
\end{align}
First, $\Pi_{\mu\nu}(Q)\vert_{\textrm{q-h}}$ and $
\Pi_{\mu\nu}(Q)\vert_{\textrm{q-a}}$ represents the quark-hole
contribution and the quark-antiquark contribution respectively
\cite{Schafer:2003jn, Rischke:2000qz}:
\begin{align}
\Pi_{\mu\nu}(Q)\vert_{\textrm{q-h}} &= m^2 \int
\frac{d\Omega}{4\pi}\hat{K_\mu}\hat{K_\nu} \left(-1 +
\frac{i\omega}{Q\cdot\hat{K}}\right), \\
\Pi_{\mu\nu}(Q)\vert_{\textrm{q-a}} &= m^2 \int
\frac{d\Omega}{4\pi}\left(\delta_{\mu4}\delta_{\nu4}+\hat{K_\mu}\hat{K_\nu}\right).
\end{align}
Second, the polarization $\delta \Pi^L(Q)$ and $\delta \Pi^T(Q)$
can be obtained as follows \cite{Le Bellac:1996}:
\begin{align}
\delta \Pi^L(Q)& =-2m^2 \frac{Q^2}{q^2}Q_1
\left(\frac{i\omega}{q}\right),\\
\delta \Pi^T(Q)& = m^2\left(\frac{i\omega}{q}\right)\left[ \left( 1
- \left(\frac{i\omega}{q}\right)^2 \right) Q_0
\left(\frac{i\omega}{q}\right) + \left(\frac{i\omega}{q}\right)
\right],
\end{align}
where $Q_0(x)=\frac{1}{2}\ln \left[(x+1)/(x-1)\right]$,
$Q_1(x)=xQ_0(x)-1$. Here one can find that the longitudinal rest
mass comes mainly  from the quark-hole contribution.

For the gluon propagator, if the external gluon momentum is soft ($Q
\simeq g\mu$), all diagram where HDL is inserted as 1PI self energy
scales in the same order  as  the  bare gluon propagator ($1/Q^2
\sim 1/ (g^2 \mu_f^2)$).  The resumed gluon propagator is given as
follows \cite{Le Bellac:1996}:
\begin{align}
{*}D^{ab}_{\mu\nu}(Q)&= \delta^{ab} \left(\frac{1}{Q^2+\delta \Pi^L
(Q)}P^L_{\mu\nu} + \frac{1}{Q^2+\delta \Pi^T
(Q)}P^T_{\mu\nu}+\frac{\xi}{Q^2}\frac{Q_\mu Q_\nu}{Q^2}\right).
\end{align}

Next we consider the quark case. If the external momenta can be
neglected against the hard scale of internal loop, the quark self
energy can be calculated as \cite{Blaizot:1993bb, Le Bellac:1996}
\begin{align}
\Sigma(P)&=-g^2 \tau^a \tau^b \int \frac{d^4 K}{(2\pi)^4} \gamma_\mu
S_F(P-K) \gamma_\nu D^{ab}_{\mu \nu} (K)
  \nonumber\\
&=m_f^2\int \frac{d \Omega}{4\pi} \frac{\slash \hspace{-0.2cm}
\hat{K}}{P \cdot \hat{K}},
\end{align}
where $P_\mu=(-\tilde{\omega},\vec{p})$ is the Euclidean quark
external momentum, $\tau^a$ is the fundamental representation of
SU(3) generator normalized as $\textrm{Tr}[\tau^a
\tau^b]=\frac{1}{2}\delta^{ab}$ and $m_f^2$ is defined as
\begin{align}
m_f^2&=\frac{1}{8}g^2C_F \left(T^2+\frac{\mu_f^2}{\pi^2}\right),
\end{align}
where $C_F=(N_c^2-1)/(2N_c)$. Again, only HDL contribution becomes
relevant at low temperature: $m_f^2 \Rightarrow \frac{1}{8}g^2C_F
\mu_f^2/\pi^2 \sim g^2 \mu_f^2$ \cite{Blaizot:1993bb}. By similar
argument with the gluon case, the resumed quark propagator can be
given as follows \cite{Le Bellac:1996}:
\begin{align}
{*}S(P)&= \frac{1}{\slash \hspace{-0.2cm} P + \Sigma(P)},
\end{align}
where
\begin{align}
\slash \hspace{-0.2cm} P + \Sigma(P) &= A_4\gamma_4 + A_s
\vec{\gamma}\cdot \hat{p},\nonumber\\
A_4 &=i\left( i\tilde{\omega} - \frac{m_f^2}{\vert \vec{p} \vert}
Q_0
\left(\frac{i\tilde{\omega}}{\vert \vec{p} \vert}\right)\right)\nonumber\\
A_s &= \vert \vec{p} \vert + \frac{m_f^2}{\vert \vec{p} \vert}\left[
1- \frac{i\tilde{\omega}}{\vert \vec{p} \vert} Q_0
\left(\frac{i\tilde{\omega}}{\vert \vec{p} \vert}\right) \right].
\end{align}

\subsubsection{Calculating the thermodynamic potential with HDL resummation }

\begin{figure}
\includegraphics[height=2.5cm]{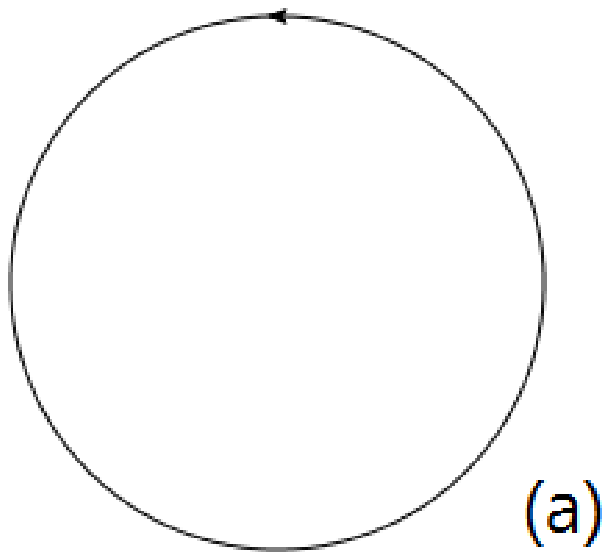}
\includegraphics[height=2.5cm]{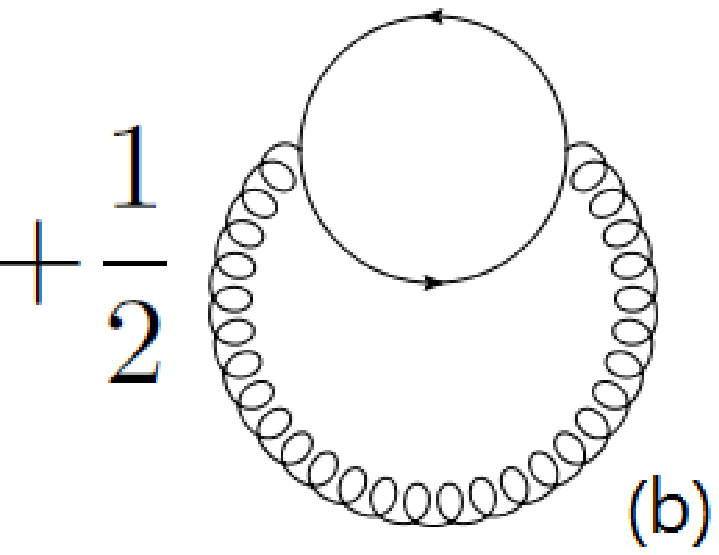}
\includegraphics[height=2.5cm]{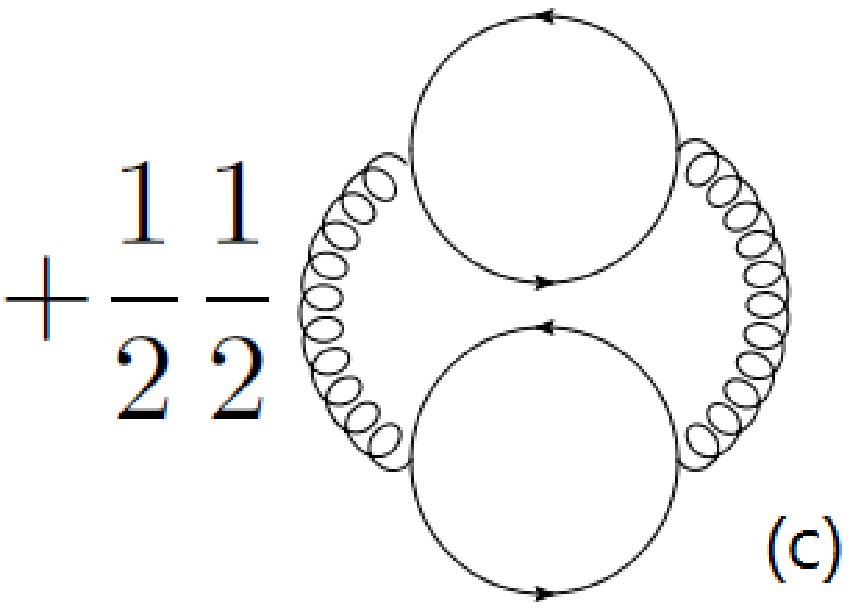}
\includegraphics[height=2.5cm]{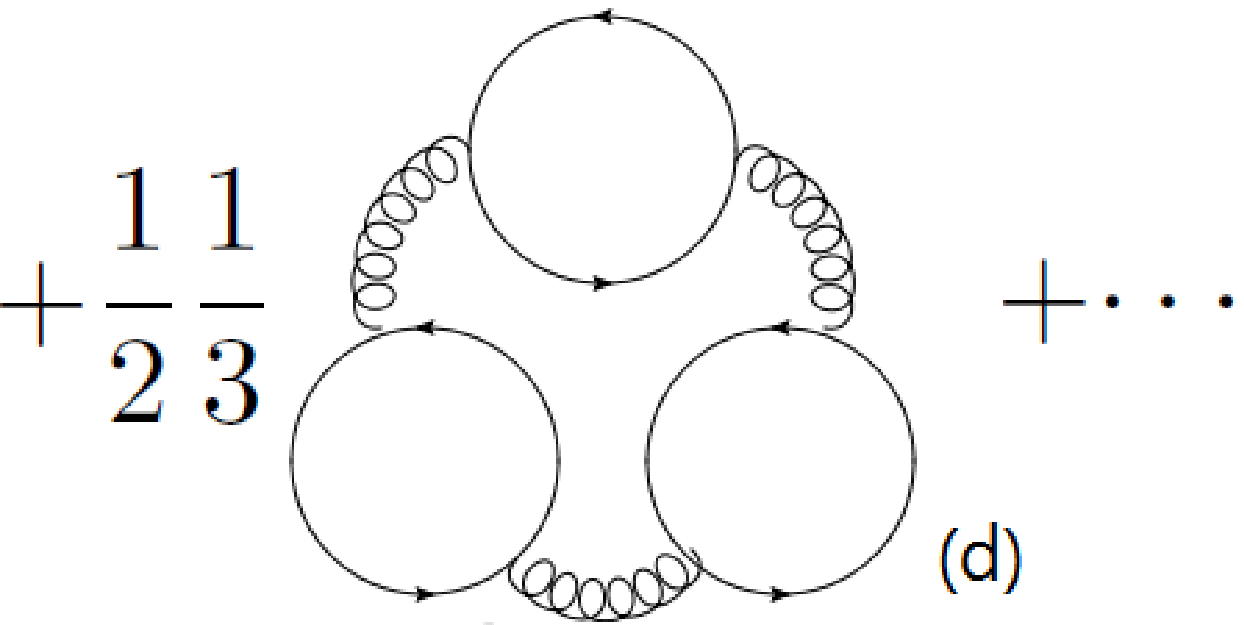}
\caption{Diagramatic description for $\ln
\mathcal{Z}_{\Omega_{q,0}}+\ln
\mathcal{Z}_{\Omega_{g,\textrm{HDL}}}$. Each diagram is presented
with its symmetric factor.}\label{diagram}
\end{figure}

The free energy can be written as
\begin{align}
\Omega(\mu)&= \langle \hat{H}\rangle-\vec{\mu}\cdot
\langle\vec{\hat{N}}\rangle=-\frac{1}{\beta }\ln \mathcal{Z}_\Omega,
\end{align}
where $\ln \mathcal{Z}_\Omega$ corresponds to the sum of all
connected ring diagrams. If the interactions are weak enough, one
can calculate elements in $\ln \mathcal{Z}_\Omega$ by perturbative
method. However, as the soft excitations in the hot/dense matter
have non-perturbative loop corrections, this quantity should be
obtained by accounting for all equivalent contribution through
resummation. In the low temperature and dense limit ($T \simeq g
\mu\ll\mu$), gluonic excitation will be mainly soft ($Q \sim T
\simeq g\mu$). In this limit, the physically relevant leading strong
interaction contribution comes from the leading quark loop
(Fig.~\ref{diagram}(a)) and from HDL resumed gluon ring diagrams
(Fig.~\ref{diagram}(b)-(d)).

The simplest diagram is Fig.~\ref{diagram}(b): a soft gluon ring
with hard quark loop correction which can also be seen as a hard
quark ring with soft gluon loop correction. As the gluon line is
soft, equivalent HDL corrections should be accounted for as a sum of
infinite series (Fig.~\ref{diagram}(c)-(d)): HDL resummation of
gluon ring diagrams.

Meanwhile, this type of resummation may not be suitable for quarks as the
physically relevant quark excitations at low temperature and dense
limit ($\vert\vec{p}\vert \sim \mu_f$) can not have HDL corrections
\cite{Schafer:2003jn}. In this limit, gluon loop should carry soft
momentum so that the correction will be suppressed by higher orders in
$\alpha_s$.
However, as there can exist suitable intermediate state for HDL
resummation among the loop integration of the quark ring trace, we
will adopt the resummation result  reported in Refs.~\cite{Baier:1999db, Andersen:2002jz}.

In this work, we consider 2-flavor light quark matter for
simplicity. Moreover near the beginning of the quark phase, the
baryon density ranges from 3-5$\rho_0$, and  the main constituent
will be the up and down quarks only. In these region, strangeness is
not the main excitation mode.  Moreover, HDL correction is not
relevant for strange quark as its mass ($m_s\sim95~\textrm{MeV}$
\cite{Agashe:2014kda}) is comparable to  $\mu_f$.

The ideal quark gas contribution (Fig.~\ref{diagram}(a)) can be
calculated as
\begin{align}
\ln \mathcal{Z}_{\Omega_{q,0}}&\simeq\beta V\left( \frac{N_c}{12}
\sum_{f=u,d} \frac{\mu_f^4}{\pi^2} \right),
\end{align}
where the absolute divergence known as ``vacuum energy'' has
been neglected. Hereafter, the matter independent  divergences will be neglected
as we are only interested in the finite difference between thermodynamic
potentials.

The ideal gluon gas contribution can be neglected ($\ln
\mathcal{Z}_{\Omega_{g,0}}\simeq (T \sim g\mu)^4$). The remaining
HDL approximated gluonic ring diagrams should be resumed. We denote
the resummation as ``the gluon resummation'' ($\ln
\mathcal{Z}_{\Omega_{g,\textrm{HDL}}}$), which corresponds to the
diagrams Fig.~\ref{diagram}(b)-(d):
\begin{align}
\ln \mathcal{Z}_{\Omega_{g,\textrm{HDL}}}&=
-\frac{(N_c^2-1)}{2}\beta V \int \frac{d^4 Q}{(2\pi)^4} \ln
\left[1+\Pi_{\mu\nu}(Q)
D_F^{~\nu\mu}(Q)\right]\nonumber\\
&= -\frac{(N_c^2-1)}{2}\beta V \int \frac{d^4 Q}{(2\pi)^4} \left(\ln
\left[1+\delta\Pi^L(Q)\frac{1}{Q^2}\right]+ 2 \ln \left[1+\delta\Pi^T(Q)\frac{1}{Q^2}\right] \right)\nonumber\\
&=\mathcal{L}+2 \mathcal{T},
\end{align}
where $\mathcal{L}$ and $\mathcal{T}$ are defined as follows:
\begin{align}
\mathcal{L} &\equiv -\frac{(N_c^2-1)}{2}\beta V \int \frac{d^4
Q}{(2\pi)^4}\ln \left[1+m^2\left( 1-\frac{i\omega}{2q}
\ln \frac{i\omega+q}{i\omega-q} \right) \frac{1}{q^2}\right], \\
\mathcal{T} &\equiv -\frac{(N_c^2-1)}{2}\beta V \int \frac{d^4
Q}{(2\pi)^4} \ln
\left[1+\frac{m^2}{2}\left(\frac{i\omega}{q}\right)\left[ \left( 1 -
\left(\frac{i\omega}{q}\right)^2 \right) Q_0
\left(\frac{i\omega}{q}\right) + \left(\frac{i\omega}{q}\right)
\right]\frac{1}{Q^2}\right].
\end{align}

Next, we take $T \rightarrow 0 $ limit and use the spatial
$d$-dimension regularization,
\begin{align}
\int \frac{d^4 Q}{(2\pi)^4} \Rightarrow \int \frac{d
\omega}{(2\pi)}\bar{\mu}^{3-d}\int \frac{d^d q}{(2\pi)^3},
\end{align}
where the discrete sum over Matsubara frequencies $T\sum_n$  changes
into a finite continuous integration, $\int d\omega / (2\pi) $, in
the  $T \rightarrow 0 $ limit. One can then  regularize the
divergence coming  from the remaining $d=D-1$ spatial dimension with
$g^2=g_0^2\bar{\mu}^{2\epsilon}=g_0^2\bar{\mu}^{3-d}$. By extracting
out the overall mass dimension as $\bar{\mu}^{3-d}$ in front of
the $d$-dimension momentum integration, one can set the coupling
constant to be dimensionless and define the one-loop counterterm in
$\ln \mathcal{Z}_{\Omega_{g,\textrm{HDL}}}$. A similar rigorous
regularization scheme in Coulomb gauge has been already reported in
Ref.~\cite{Andersen:1999sf}. In the present work, the calculation is
performed in the covariant gauge and the result is totally
equivalent to the calculation of Ref.~\cite{Andersen:1999sf,
Andersen:2002jz} at $T=0$.

As both $\delta \Pi^L(Q)$ and $\delta \Pi^L(Q)$ depend only on
$i\omega/q$, the integration variable $\omega$ can be re-scaled as
$\omega\rightarrow q\bar{\omega}$ to scale out the dimensions of the integration:
\begin{align}
\mathcal{L} &= -\frac{(N_c^2-1)}{2}\beta V
\frac{1}{(2\pi)}\frac{d\Omega_d}{(2\pi)^d}\bar{\mu}^{3-d}~2\int_0^\infty
d \bar{\omega} \int_0^\infty dqq^d \ln
\left[q^2+\widetilde{\mathcal{L}}(\bar{\omega})\right],
\end{align}
where the absolute divergence proportional to $\int_0^\infty dq q^d
\ln q^2$ has been neglected and
$\widetilde{\mathcal{L}}(\bar{\omega})$ is given as
\begin{align}
\widetilde{\mathcal{L}}(\bar{\omega}) & \equiv m^2\left(
1-\frac{i\bar{\omega}}{2} \ln
\frac{i\bar{\omega}+1}{i\bar{\omega}-1} \right).
\end{align}
Using the following integration formula \cite{Andersen:1999sf}
\begin{align}
\int_0^\infty dkk^\alpha \ln(k^2+m^2)&=\frac{
\Gamma\left(\frac{1+\alpha}{2}\right)\Gamma\left(\frac{1-\alpha}{2}\right)}{\alpha+1}m^{\alpha+1}+\textrm{absolute
divergence},
\end{align}
and the $\overline{\textrm{MS}}$ subtraction scheme
$\mu_4^2=(4\pi/e^\gamma)\bar{\mu}^2$, the integration can be
arranged as
\begin{align}
\mathcal{L} &= (N_c^2-1)\beta V
\frac{1}{(2\pi)}\frac{d\Omega_3}{(2\pi)^3}\frac{(m^2)^2}{4}\left[\left(1-\ln\frac{m^2}{\pi
\mu_4^2}\right)\alpha-\beta+\frac{1}{\epsilon}\alpha\right]\nonumber\\
& \Rightarrow \beta V \left[ \alpha_s^2
\frac{2}{\pi}\left(\sum_{f=u,d}\frac{\mu_f^2}{\pi^2}\right)^2
\left[\left(1- \ln2  -
\ln\left(\sum_{f=u,d}\frac{\mu_f^2}{\pi^2}\frac{1}{\mu_4^2}\right)
-\ln \alpha_s\right)\alpha - \beta\right] \right]_{\textrm{finite}},
\end{align}
where $N_c=3$ and the constants are calculated as
\begin{align}
l(\bar{\omega}) &\equiv 1+\bar{\omega} \left( \arctan {\bar{\omega}}
-\frac{\pi}{2}\right),\\
\alpha&\equiv\int_0^\infty
d\bar{\omega}l(\bar{\omega})^2=\frac{1}{3}\pi(1-\ln2)=0.321336
,\\
\beta&\equiv\int_0^\infty d \bar{\omega}l(\bar{\omega})^2 \ln
l(\bar{\omega})=-0.176945 .
\end{align}

$\mathcal{T}$ can be obtained in a similar way. After using the same
re-scaling $\omega\rightarrow q\bar{\omega}$, the integration can be
written as
\begin{align}
\mathcal{T} &= -\frac{(N_c^2-1)}{2}\beta V
\frac{1}{(2\pi)}\frac{d\Omega_d}{(2\pi)^d}\bar{\mu}^{3-d}
2\int_0^\infty d \bar{\omega} \int_0^\infty dq~q^d \ln
\left[q^2+\widetilde{\mathcal{T}}(\bar{\omega})\right],
\end{align}
where $\widetilde{\mathcal{T}}(\bar{\omega})$ is given as
\begin{align}
\widetilde{\mathcal{T}}(\bar{\omega}) & \equiv \frac{m^2}{2}\left(
-\frac{\bar{\omega}^2}{\bar{\omega}^2+1}+\frac{i\bar{\omega}}{2} \ln
\frac{i\bar{\omega}+1}{i\bar{\omega}-1} \right).
\end{align}
Again, using similar method, the integration can be arranged as
\begin{align}
\mathcal{T} &= (N_c^2-1)\beta V
\frac{1}{(2\pi)}\frac{d\Omega_3}{(2\pi)^3}\frac{(m^2)^2}{8}\left[\left(1-\ln\frac{m^2}{2\pi
\mu_4^2}\right)\frac{1}{2}\bar{\alpha}-\frac{1}{2}\bar{\beta}+\frac{1}{2\epsilon}\bar{\alpha}\right]\nonumber\\
& \Rightarrow \beta V \left[ \alpha_s^2
\frac{1}{\pi}\left(\sum_{f=u,d}\frac{\mu_f^2}{\pi^2}\right)^2
\left[\left(1 -
\ln\left(\sum_{f=u,d}\frac{\mu_f^2}{\pi^2}\frac{1}{\mu_4^2}\right)
-\ln \alpha_s
\right)\frac{1}{2}\bar{\alpha}-\frac{1}{2}\bar{\beta}\right]
\right]_{\textrm{finite}},
\end{align}
where $N_c=3$ and the constants are calculated as
\begin{align}
t(\bar{\omega}) &\equiv
-\frac{\bar{\omega}^2}{\bar{\omega}^2+1}+\bar{\omega}
\left(\frac{\pi}{2} -\arctan {\bar{\omega}}\right),\\
\bar{\alpha}&\equiv\int_0^\infty d
\bar{\omega}t(\bar{\omega})^2=\frac{1}{12}\pi(-5+8\ln2)=0.142727
,\\
\bar{\beta}&\equiv \int_0^\infty d \bar{\omega}t(\bar{\omega})^2 \ln
t(\bar{\omega})=-0.200869.
\end{align}
The overall divergence has been subtracted by the following
counterterm ($\epsilon \rightarrow 0 $)
\begin{align}
\Delta \ln \mathcal{Z}_{\Omega}&=-(3^2-1)\beta V
\frac{1}{(2\pi)}\frac{d\Omega_3}{(2\pi)^3}\frac{(m^2)^2}{4}\left(\alpha+\frac{1}{2}\bar{\alpha}\right)\frac{1}{\epsilon}
.
\end{align}

\begin{figure}
\includegraphics[height=2.5cm]{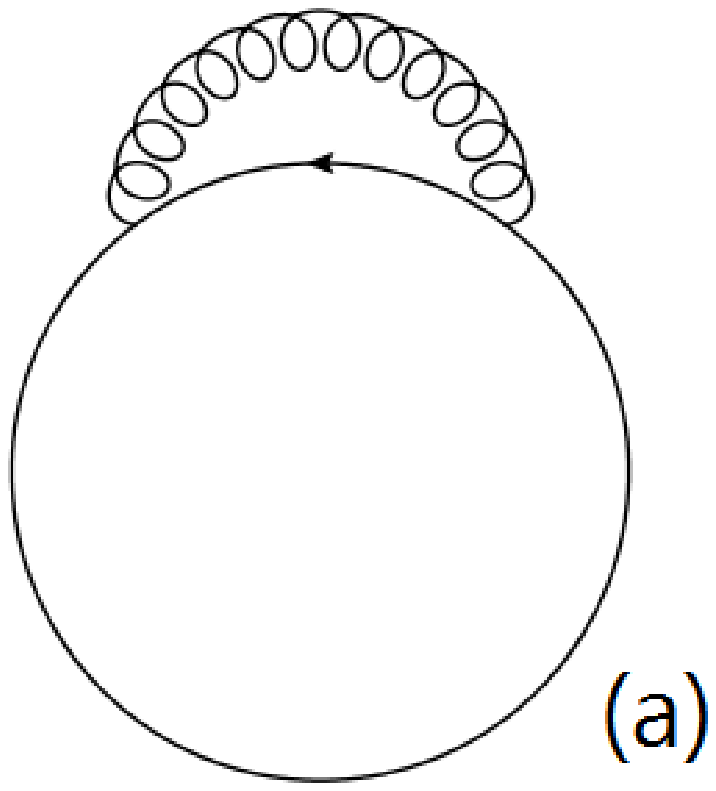}
\includegraphics[height=2.5cm]{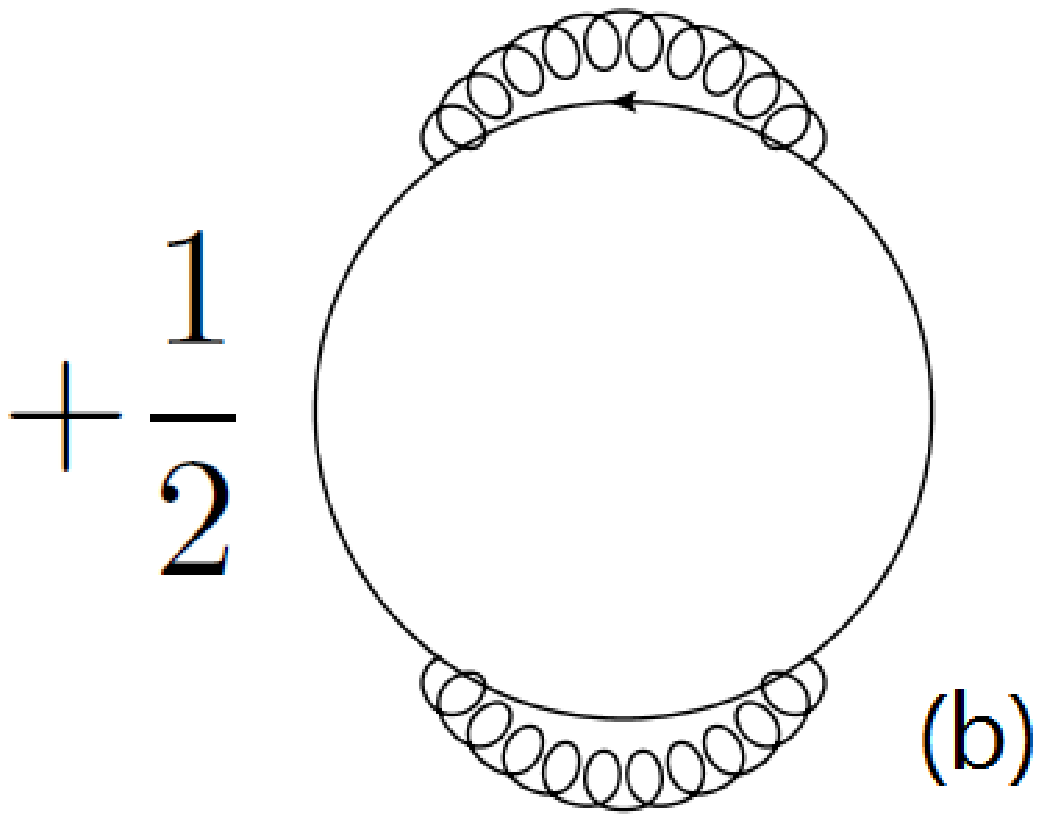}
\includegraphics[height=2.5cm]{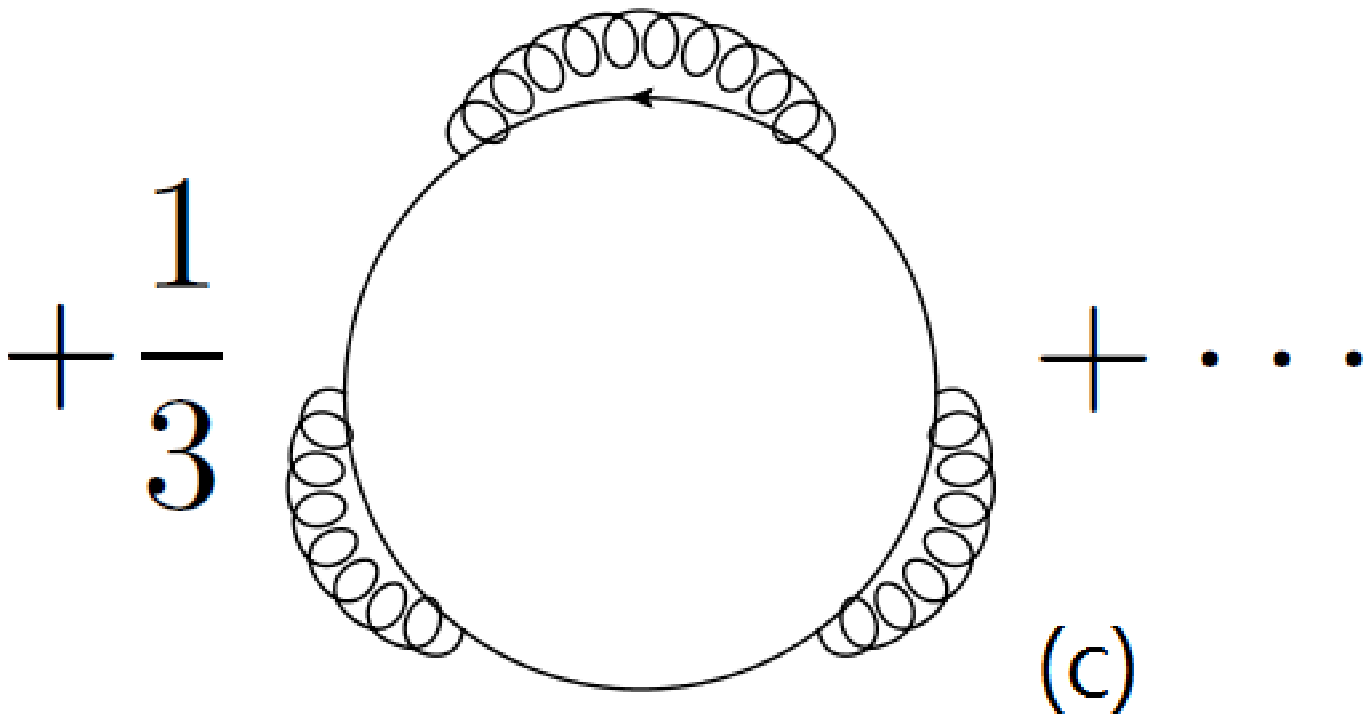}
\caption{Diagramatic description for $\ln
\mathcal{Z}_{\Omega_{q,\textrm{HDL}}}$. Each diagram is presented
with its symmetric factor.}\label{diagram2}
\end{figure}

For the quarks, the resummation can be done in a similar way as in the
gluon case \cite{Andersen:2002jz}. We adopt the result from
Ref.~\cite{Andersen:2002jz} and denote it as ``the quark
resummation'' ($\ln\mathcal{Z}_{\Omega_{q,\textrm{HDL}}}$), which
corresponds to the diagrams in Fig.~\ref{diagram2}:
\begin{align}
\ln \mathcal{Z}_{\Omega_{q,\textrm{HDL}}}&=\beta V\frac{1}{4}
\sum_{f=u,d}\frac{\mu_f^4}{\pi^2}\left[-4\left(\frac{\alpha_s}{\pi}\right)+
\left(\frac{8}{3}-\frac{4}{9}\pi^2\right)\left(\frac{\alpha_s}{\pi}\right)^2\right].
\end{align}

Finally,  $\ln \mathcal{Z}_{\Omega}=\ln \mathcal{Z}_{\Omega_{q,0}}+\ln
\mathcal{Z}_{\Omega_{g,\textrm{HDL}}}+\ln
\mathcal{Z}_{\Omega_{q,\textrm{HDL}}}$ can be written as an expansion in $\alpha_s^n(\ln\alpha_s)^m$ as follows:
\begin{align}
\ln \mathcal{Z}_{\Omega}=&~\beta V
\Biggl(\frac{1}{4}\sum_{f=u,d}\frac{\mu_f^4}{\pi^2}
\left[1-4\left(\frac{\alpha_s}{\pi}\right)
+\left(\frac{8}{3}-\frac{4}{9}\pi^2\right)\left(\frac{\alpha_s}{\pi}\right)^2\right]\nonumber\\
&+\alpha_s^2~\frac{2}{\pi}\left(\sum_{f=u,d}\frac{\mu_f^2}{\pi^2}\right)^2
\left[\left(1-\ln \alpha_s-
\ln\left(\sum_{f=u,d}\frac{\mu_f^2}{\pi^2}\frac{1}{\mu_4^2}\right)\right)\Lambda_1-\Lambda_2-\alpha
\ln2 \right]\Biggr),
\end{align}
where the constants  $\Lambda_1 \equiv
\alpha+\frac{1}{2}\bar{\alpha}$ and $\Lambda_2 \equiv
\beta+\frac{1}{2}\bar{\beta}$. The thermodynamic quantities can be
obtained from $\ln \mathcal{Z}_{\Omega}$:
\begin{align}\allowdisplaybreaks
\frac{\Omega(\mu)}{V}=& \frac{\langle \hat{H}\rangle-\vec{\mu}\cdot
\langle\vec{\hat{N}}\rangle}{V}=-\frac{1}{\beta V}\ln
\mathcal{Z}_\Omega\nonumber\\
=&-\frac{1}{4}\sum_{f=u,d}\frac{\mu_f^4}{\pi^2}\left[1-4\left(\frac{\alpha_s}{\pi}\right)
+\left(\frac{8}{3}-\frac{4}{9}\pi^2\right)\left(\frac{\alpha_s}{\pi}\right)^2\right]\nonumber\\
&-\alpha_s^2\frac{2}{\pi}\left(\sum_{f=u,d}\frac{\mu_f^2}{\pi^2}\right)^2
\left[\left(1-\ln \alpha_s-
\ln\left(\sum_{f=u,d}\frac{\mu_f^2}{\pi^2}\frac{1}{\mu_4^2}\right)\right)\Lambda_1-\Lambda_2-\alpha
\ln2 \right],\label{gpot}\\
\rho_i(\mu)=&\frac{\langle \hat{N}_i \rangle}{V}=\frac{1}{\beta V}
\frac{\partial}{\partial \mu_i}\ln
\mathcal{Z}_\Omega,\nonumber\\
=&~\frac{\mu_i^3}{\pi^2}\left[1-4\left(\frac{\alpha_s}{\pi}\right)
+\left(\frac{8}{3}-\frac{4}{9}\pi^2\right)\left(\frac{\alpha_s}{\pi}\right)^2\right]\nonumber\\
&+\alpha_s^2\frac{8}{\pi}
\left(\sum_{f=u,d}\frac{\mu_f^2}{\pi^2}\right)
\frac{\mu_i}{\pi^2}\left[\left(\frac{1}{2}-\ln \alpha_s-
\ln\left(\sum_{f=u,d}\frac{\mu_f^2}{\pi^2}\frac{1}{\mu_4^2}\right)
\right)\Lambda_1-\Lambda_2-\alpha
\ln2 \right],\label{nd}\\
\epsilon(\mu)=&\frac{\langle \hat{H} \rangle}{V}=-\frac{1}{V}\left(
\frac{\partial}{\partial\beta}-\frac{1}{\beta} \vec{\mu}\cdot
\frac{\partial}{\partial \vec{\mu}}\right) \ln
\mathcal{Z}_\Omega\nonumber\\
=&~\frac{3}{4}
\sum_{f=u,d}\frac{\mu_f^4}{\pi^2}\left[1-4\left(\frac{\alpha_s}{\pi}\right)
+\left(\frac{8}{3}-\frac{4}{9}\pi^2\right)\left(\frac{\alpha_s}{\pi}\right)^2\right]\nonumber\\
&+\alpha_s^2
\frac{6}{\pi}\left(\sum_{f=u,d}\frac{\mu_f^2}{\pi^2}\right)^2
\left[\left(\frac{1}{3}-\ln \alpha_s -
\ln\left(\sum_{f=u,d}\frac{\mu_f^2}{\pi^2}\frac{1}{\mu_4^2}\right)\right)\Lambda_1-\Lambda_2-\alpha
\ln2\right]\label{qed},
\end{align}
where $\alpha_s\left(\mu_4 \right)$ is assigned for 2-loop
renomalization in $\overline{\textrm{MS}}$ scheme:
\begin{align}
\alpha_s\left(\mu_4 \right) & = \frac{4\pi}{\beta_0
\ln\left(\mu_4^2/\Lambda^2_{\overline{\textrm{MS}}}\right)
}\left(1-\frac{2 \beta_1 \ln \left(
\ln\left(\mu_4^2/\Lambda^2_{\overline{\textrm{MS}}}\right)\right)} {\beta_0^2 \ln\left(\mu_4^2/\Lambda^2_{\overline{\textrm{MS}}}\right)}\right),\nonumber\\
\beta_0 & = (11N_c-2N_f)/3,\nonumber\\
\beta_1 & = (34N_c^2-13N_c N_f+3N_f/N_c)/6,
\end{align}
where $\mu_4$ will be parameterized as $1.5\mu \leq \mu_4 \leq
4\mu$. $\mu$ is the quark chemical potential at iso-spin symmetric
condition. As discussed in Ref.~\cite{Karsch:1997gj, Baier:1999db,
Andersen:2002jz}, if higher order calculation can be obtained,
$\alpha_s$ can be determined by solving the self-consistent gap
equation for the thermodynamic potential. On the other hand, since
we have calculated only to leading order in the HDL perturbation,
the effect of $\alpha_s\left(\mu_4 \right)$ is investigated by
varying $ 1.5 \mu\leq \mu_4 \leq 4\mu$.

\begin{figure}
\includegraphics[width=8cm]{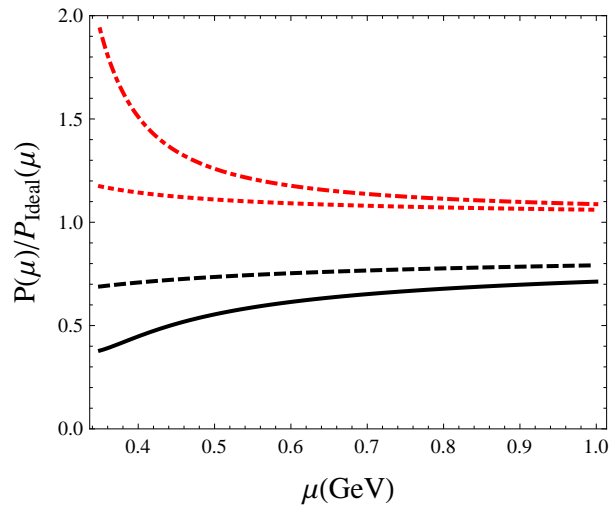}
\caption{(Color online) The ratio $P(\mu)/P_{\textrm{ideal}}(\mu)$
with $\mu_4=2\mu $ for red dot-dashed  and black solid line, and
$\mu_4=4\mu$ for red dotted and  black dashed line. The upper two
lines are obtained with gluon resummation results and the lower two
lines from adding the quark resummation to the gluon
resummation.}\label{pratio}
\end{figure}

\begin{figure}
\includegraphics[width=8cm]{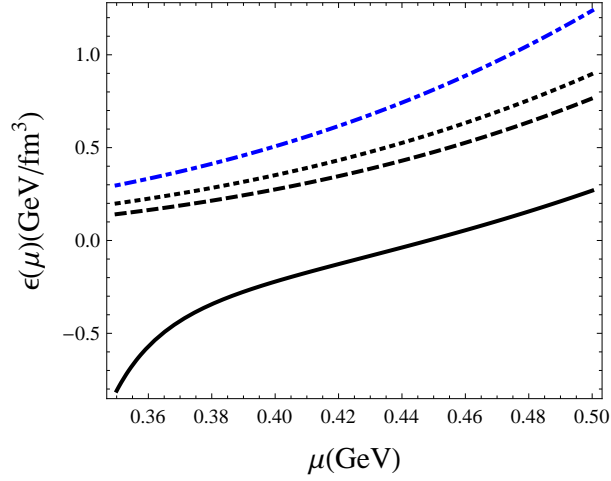}
\caption{(Color online) Quark energy density $\epsilon(\mu)$ with
various $\mu_4$. Blue dot-dashed line is for ideal quark gas. Black
lines represent the result  for quark and gluon resummation. Black
dotted line represents the result with $\mu_4=4\mu$, black dashed
line with $\mu_4=2.5\mu$ and black solid line with $\mu_4=1.5\mu$,
respectively.}\label{edensity}
\end{figure}

Fig.\ref{pratio} shows the ratio $P(\mu)/P_{\textrm{ideal}}(\mu)$
with various $\mu_4$. The upper two lines are obtained with
contributions coming from  gluon HDL resummation results and the
lower two lines from adding the quark resummation to the gluon HDL
resummation. Red dotdashed  and black solid line are obtained with
$\mu_4=2\mu$. Red dotted and black dashed line are obtained with
$\mu_4=4\mu$. One can find that HDL resumed
$P(\mu)/P_{\textrm{ideal}}(\mu)$ varies in a large band in
Fig~\ref{pratio}. As can be expected,
$P(\mu)/P_{\textrm{ideal}}(\mu)$ approaches to 1 as $\mu_4$ becomes
larger.  Here one notes that the gluon resummation gives the
enhanced ratio ($P(\mu)/P_{\textrm{ideal}}(\mu)\sim 1.1$ at
$\mu=1~\textrm{GeV}$), while inclusion of the quark resummation
gives the reduced ratio ($P(\mu)/P_{\textrm{ideal}}(\mu)\sim 0.8$ at
$\mu=1~\textrm{GeV}$). The reduction of pressure due to the
inclusion of quark contributions is consistent with previous studies
that include a massive strangeness and the quark interaction part up
to $O(\alpha_s^2)$ \cite{Kurkela:2009gj, Fraga:2013qra}. Based on
the result of Ref.~\cite{Kurkela:2009gj},  the authors in
Ref.~\cite{Fraga:2013qra} also find  a similar behavior of the pQCD
ratio ($0.6 <P(\mu)/P_{\textrm{ideal}}(\mu)< 0.8$ near
$\mu=1~\textrm{GeV}$) as shown in Fig.~2 of their work. While the
quark resummation enhances the $O(\alpha_s)$ correction compared to
the weak coupling expansion \cite{Baier:1999db, Andersen:2002jz},
introducing the quark phase in the core of a neutron star in general
allows for a more massive neutron star that is consistent with
recent measurement as was discussed in Ref.~\cite{Fraga:2013qra}. To
fix the scale, we plotted the quark energy density in
Fig.~\ref{edensity}. One notes that the value becomes negative at
lower $\mu$ when $\mu_4 = 1.5 \mu$. This behavior is due to the
increase in $\alpha_s$ at small scale, which is a consequence of
asymptotic freedom. To make sure the quark energy density be
positive, we choose $\mu_4 = 2.5 \mu$ throughout this section.

From Eqs.~\eqref{gpot}-\eqref{qed}, one can find that the
thermodynamic quantities depend mainly on quark and gluon rest mass
$m^2,m_f^2\sim g^2 \mu^2/\pi^2$. In the HDL resummation, the gluonic
contribution is dominated by the longitudinal gluon rest mass.
Iso-spin asymmetric properties can be accounted for by assigning
asymmetric quark chemical potential to the quark and gluon rest
mass.

\subsection{Cold dense matter symmetry energy at normal phase}

\begin{figure}
\includegraphics[width=8cm]{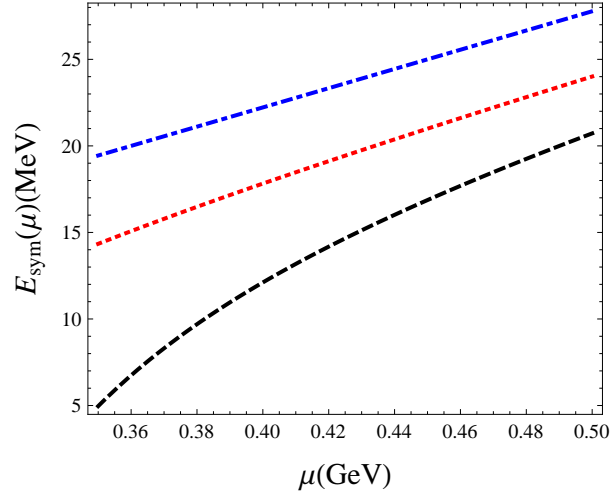}
\caption{(Color online) Quark matter symmetry energy at normal phase
($\mu_4=2.5\mu$). Blue dot-dashed line represents symmetry energy of
ideal quark gas. Red dotted line represents symmetry energy with
gluon resummation only while black dashed line represents the result
after adding the quark resummation.}\label{symen}
\end{figure}

In the iso-spin asymmetric matter, the light quark chemical
potential can be defined as follows:
\begin{align}
\mu^u_d=\mu \left(1\mp \frac{1}{3}I_B\right)
^\frac{1}{3}.\label{asymdef1}
\end{align}
By assigning this relation to Eqs.~\eqref{nd}-\eqref{qed}, quark
matter symmetry energy in the normal phase can be obtained from
energy per baryon number as follows:
\begin{align}
\frac{\epsilon(\mu,I_B)}{\rho_B(\mu,I_B)} &= \bar{E}(\mu,I_B) = E_0(\mu,I_B)+\bar{E}_{sym}(\mu)I^2_B + O(I_B^4) +\cdots,\label{epb}\\
E_{sym}(\mu)&= \frac{1}{2!} \frac{\partial^2}{\partial
I_B^2}\bar{E}(\mu,I_B)\label{qmsym},
\end{align}
where $\rho_B=(\rho_u+\rho_d)/3$, $I_B=
3(\rho_u-\rho_d)/(\rho_u+\rho_d)$.

The quark matter symmetry energy is plotted in Fig.~\ref{symen}.
Here one can check that the symmetry energy is reduced by HDL
resummation compared to the ideal quark gas case. It reduces
slightly when the gluon HDL resummation is included and more when
the quark resummation is added. Hence, one can conclude that HDL
resumed interaction makes it easier for the cold quark matter to be
iso-spin asymmetric: it costs less energy to reach the same iso-spin
asymmetry in comparison with the ideal quark gas. When temperature
becomes even lower, additional non-perturbative effect should be
considered. We discuss an effect in the next section.

\section{Symmetry energy at extremely low temperature}
Now we consider the extremely cold matter case ($T \ll g \mu$). As
we have found in the previous section, the gluonic contribution to
symmetry energy depends mainly on the iso-spin asymmetric quark and
gluon rest masses which comes from HDL resummation. In this section
we will concentrate on the iso-spin asymmetric corrections which
come from extremely cold matter.

In such a cold condition, for ideal quark gas, all fermion will be
confined in their Fermi sea without large fluctuation. When
interaction turns on, one can consider nontrivial correlation
between fermions and related effect. According to effective
descriptions based on Wilson's renormalization group approach
\cite{Wilson:1971bg, Wilson:1971dh, Polchinski:1992ed,
Shankar:1993pf}, when one scales the momentum to near the Fermi
surface, the interaction between fermions with opposite momenta
becomes marginal. So if the interaction is attractive, it is natural
to form a bosonic condensate for two fermions on the same Fermi sea with
opposite Fermi velocity (BCS pairing) \cite{Bardeen:1957mv}. In QCD,
one-gluon exchange via color antisymmetric channel naturally
provides attractive interaction. So in the extremely low temperature
limit, one should consider the effects of color superconductivity
\cite{Alford:1997zt, Alford:2007xm}.

\subsection{2-color superconductivity (2SC) and symmetry energy}

\begin{figure}
\includegraphics[width=8cm]{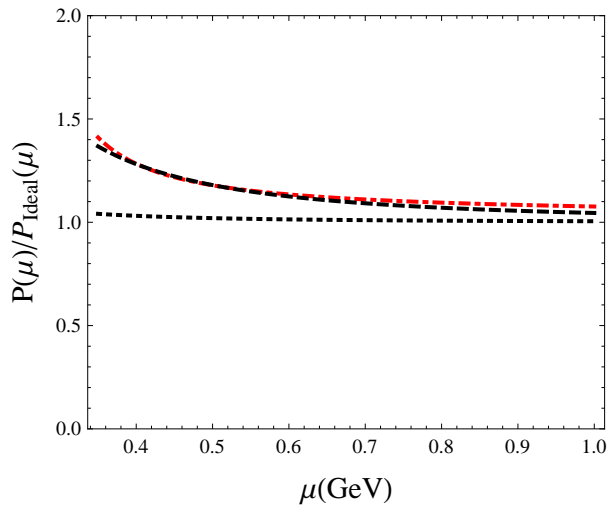}
\caption{(Color online)  Ratio $P(\mu)/P_{\textrm{ideal}}(\mu)$ with
various $\Delta$. Black dashed line represents
$P(\mu)/P_{\textrm{ideal}}(\mu)$ with $\Delta=150~\textrm{MeV}$.
Black dotted line represents $P(\mu)/P_{\textrm{ideal}}(\mu)$ with
$\Delta=50~\textrm{MeV}$. Red dot-dashed line represents
$P(\mu)/P_{\textrm{ideal}}(\mu)$ of the gluon resummation with
$\mu_4=2.5\mu$ at normal phase as a criterion.}\label{pratio2sc}
\end{figure}

First, we confine the density region to be similar to that of
previous section: 3-5$\rho_0$. The corresponding quark chemical
potential then  ranges from $0.35~\textrm{GeV} \leq \mu \leq
0.5~\textrm{GeV}$. In this region, there is a mismatch of Fermi
momentum between light quark and strange quark ($ p_F^u \simeq
p_F^d\simeq\mu \gg p_F^s=\sqrt{\mu^2_s-m^2_s}$) even if same numbers
of light quarks and strange quarks exist. Thus when strangeness is
not the main excitation, only light quark flavor can participate in
formation of BCS condensate (2SC) \cite{Rajagopal:2000wf}. So again,
we consider only light quark flavors in this section. The spin-0
channel forms the most stable diquark condensate as quark pairs from
the whole iso-tropic Fermi surface can contribute to the pairing.
This type of condensation requires the quark pair to be on the same
Fermi surface but with opposite velocity and spin alignment, such
that they are in same helicity state. As color configuration is
already antisymmetric, to obtain total antisymmetric wave function,
quark flavors should be in the antisymmetric configuration. The
diquark condensate can then be summarized as follows
\cite{Casalbuoni:2001ha, Nardulli:2002ma}:
\begin{align}
\left \langle \psi^T_{L,\alpha i} C \psi_{L,\beta j} \right\rangle
=-\left \langle \psi^T_{R,\alpha i} C \psi_{R,\beta j} \right\rangle
= \frac{\Delta}{2} \epsilon_{\alpha \beta 3}\epsilon_{i j 3},
\end{align}
where $C = i\sigma_2$, $\Delta$ is the superconducting gap size, the
Greek index represents color, the Latin index represents quark
flavor and $L$, $R$ represents chirality of the quarks. The condensate
can be included in the interaction terms of Lagrangian as the
invariant coupling \cite{Casalbuoni:2001ha, Nardulli:2002ma}:
\begin{align}
\mathcal{L}_\Delta = -\frac{\Delta}{2} \psi^T_{L} C \epsilon
\psi_{L} \epsilon - (L \rightarrow R) + \textrm{h.c.},\label{2scil}
\end{align}
where $\epsilon=i\sigma_2$ on SU(2) color-flavor index.

Although the existence of the 2SC phase is based on the  iso-spin
symmetric Fermi sea of the light quarks at zero temperature, the
phase can persist up to some critical temperature and iso-spin
asymmetry determined by the  Clogston limit $\delta \mu =
\Delta/\sqrt{2}$ \cite{Alford:2007xm, Clogston:1962zz}.
Specifically, the PNJL model calculation \cite{Roessner:2006xn}
suggests that the critical temperature is in the order of 50 MeV.
Moreover, even if the  pure neutron matter ($I_B=1$) changes into
quark matter, the difference between Fermi sea of $u$ and $d$ quark
will be $\delta \mu \sim 0.23 \mu$. Hence, assuming that the gap
also grows with the chemical potential, the Clogston limit can be
satisfied at higher density. In realistic  heavy ion collisions
involving ordinary dense matter ($I_B<1$) the iso-spin difference
will be smaller ($\delta \mu < 0.23 \mu$) and and the possibility to
probe 2SC matter remains. Such a matter can be produced in the
planed FAIR experiment \cite{Stocker:2015cva}.

If one considers only the ideal quark contribution and the
correction coming from the non-perturbative gap contribution, the
free energy can be written as follows \cite{Schafer:1999fe,
Kapusta:2006}:
\begin{align}
\Omega_{\Delta}(\mu) & = \Omega_{q,0}(\mu) +
\delta\Omega_{q,\Delta}(\mu)\nonumber\\
& =-\frac{1}{12}\sum^{N_c}\sum_{f=u,d}\frac{\mu_f^4}{\pi^2} +
\sum_{i}^{2\textrm{SC}}\frac{\mu_i^2}{\pi^2}\int
dp_4\left[\sqrt{p^2_4+\Delta(p_4)^2}-p_4-\frac{\Delta(p_4)^2}{2\sqrt{p^2_4+\Delta(p_4)^2}}\right]\nonumber\\
& \simeq -\frac{1}{12}\sum^{N_c}\sum_{f=u,d}\frac{\mu_f^4}{\pi^2}
-\sum_{i}^{2\textrm{SC}}\frac{\mu_i^2 \Delta^2}{4\pi^2},\label{g2sc}
\end{align}
where the approximation in the last line follows from
Ref.~\cite{Alford:2007xm, Miransky:1999tr}. In the 2SC phase, the
2-color and 2-flavor states participate in the condensation and
hence share the same chemical potential. Here, we ignore $u$ and $d$
quark chemical potential asymmetry coming from the difference in
their charge. Then, the correction from 2SC can be written as
$\delta\Omega_{q,\Delta}(\mu)=-4(\mu^2 \Delta^2/4\pi^2)$ without the
iso-spin index.

The gap size $\Delta$ has been determined in several studies. The
first work in non-perturbative regime was reported by D. T. Son as
$\Delta / \mu= (b/g^5) \textrm{exp}\left(-3\pi^2/\sqrt{2}g\right)$
by solving the self-consistent gap equation in using HDL
approximation \cite{Son:1998uk}. Sch$\ddot{a}$fer and Wilczek
\cite{Schafer:1999jg} determined the constant to be $b=256\pi^4$ and
the corresponding gap size in $\mu=400~\textrm{MeV}$ to be in the
order of 100 MeV. The estimation from HDET \cite{Nardulli:2002ma}
was found to be in the order of 50 MeV, while model calculation
\cite{Huang:2003xd} gives the value in the order of 100 MeV.

In Fig.~\ref{pratio2sc} we plotted $P(\mu)/P_{\textrm{ideal}}(\mu)$
in the range  of $50~\textrm{MeV} \leq \Delta \leq
200~\textrm{MeV}$. As one can expect, as $\Delta$ becomes larger,
the pressure deviates more from the ideal quark gas limit. When
$\Delta=150~\textrm{MeV}$, $P(\mu)/P_{\textrm{ideal}}(\mu)$ shows
similar behavior to $P(\mu)/P_{\textrm{ideal}}(\mu)$ at normal phase
(with only the gluon resummation, $\mu_4=2.5\mu$). Also with the gap
size $\Delta=150~\textrm{MeV}$, large enough Clogston limit ($\delta
\mu \sim 0.23 \mu < \Delta/\sqrt{2}$) can be obtained in 3-5$\rho_0$
region. For gluonic terms, at 2SC phase, HDL effective Lagrangian
can not be inserted directly to the present Lagrangian as the
quasi-loop correction in 2SC phase is different with the HDL of
normal phase. However, it was argued in Ref.~\cite{Rischke:2000cn}
that the coupling in such cases will be weak, hence, may give
minimal effect to the free energy. In this section, we choose the
$\Delta=150~\textrm{MeV}$ to match the value of the free energy
calculated in both phases as shown in the upper two lines in
Fig.~\ref{pratio2sc}.

\begin{figure}
\includegraphics[width=8cm]{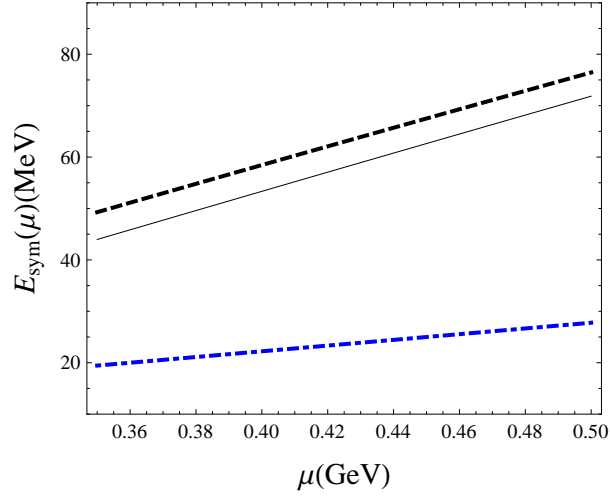}
\caption{(Color online) Quark matter symmetry energy at 2SC phase.
Blue dot-dashed line represents symmetry energy of the ideal quark
gas. Black solid line represents symmetry energy at 2SC phase with
$\Delta=200~\textrm{MeV}$. Black dashed line represents symmetry
energy at 2SC phase with $\Delta=150~\textrm{MeV}$.}\label{syme2sc}
\end{figure}

The symmetry energy in 2SC phase can be obtained in similar way from
the energy per quasi-baryon number:
\begin{align}
\frac{\epsilon(\mu,I_{\tilde{B}})}{\rho_{\tilde{B}}(\mu,I_{\tilde{B}})} & =  \bar{E}(\mu,I_{\tilde{B}})
= E_0(\mu,I_{\tilde{B}})+\bar{E}_{sym}(\mu)I^2_{\tilde{B}} + O(I_{\tilde{B}}^4) +\cdots,\label{epqb}\\
E^{\textrm{2SC}}_{sym}(\mu)&= \frac{1}{2!}
\frac{\partial^2}{\partial
I_{\tilde{B}}^2}\bar{E}(\mu,I_{\tilde{B}})\label{qmsym2sc},
\end{align}
where
$\rho_{\tilde{B}}=(\rho_{\textrm{unpaired}}+\rho_{\textrm{paired}})/3$
and $I_{\tilde{B}}=I_B/3 $ as asymmetrizable unpaired states reduce
to $1/3$ of the normal phase. From $\Omega_{\Delta}(\mu)$, the
quasi-quark number density and the energy density can be obtained as
follows:
\begin{align}
\rho_i(\mu)&=\frac{1}{3}\frac{\mu_i^3}{\pi^2},\\
\rho_{\Delta i}(\mu)&=\frac{1}{3}\frac{\mu_i^3}{\pi^2}+\frac{\mu_i
\Delta^2}{2\pi^2},\\
\epsilon_\Delta (\mu) & =\epsilon_{\textrm{unpaired}}(\mu)+
\epsilon_{\textrm{paired}}(\mu) \nonumber\\
& =\frac{1}{4}
\sum^{\textrm{unpaired}}_i\frac{\mu_i^4}{\pi^2}+\frac{1}{4}
\sum^{2\textrm{SC}}_i\left[\frac{\mu_i^4}{\pi^2}+ \frac{\mu_i^2
\Delta^2}{ \pi^2}\right],
\end{align}
where $\mu^u_d=\mu \left(1\mp I_{\tilde{B}}\right) ^\frac{1}{3}$ is
assigned only for the unpaired states, while the other  quasi-quark
states share symmetric chemical potential $\mu$. The symmetry energy
in 2SC phase is plotted in Fig.~\ref{syme2sc}. It is substantially
larger than the symmetry energy in the normal phase: it costs more
energy to reach the same iso-spin asymmetry in comparison with the
normal phase. This large difference may lead to quite different
phenomenological prediction, which will be discussed in last
section. As the gluonic coupling will be weak \cite{Rischke:2000cn},
the result in Fig.~\ref{syme2sc} could be dominant contribution.

\subsection{Gluon rest masses in iso-spin asymmetric 2SC matter}

\begin{figure}
\includegraphics[height=2.8cm]{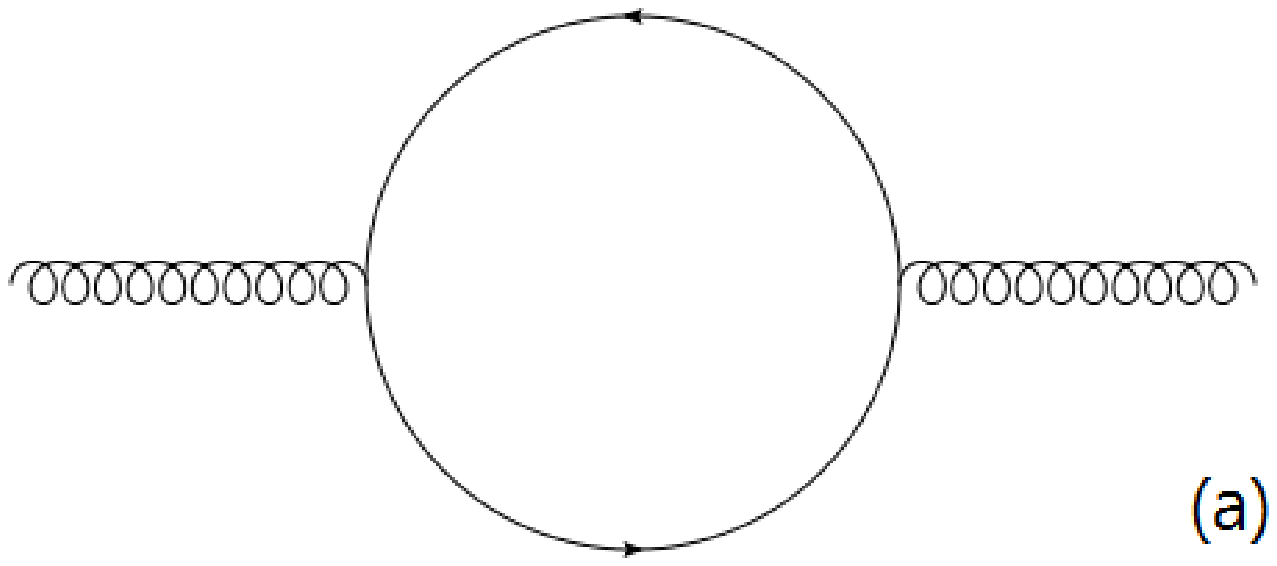}
\includegraphics[height=2.8cm]{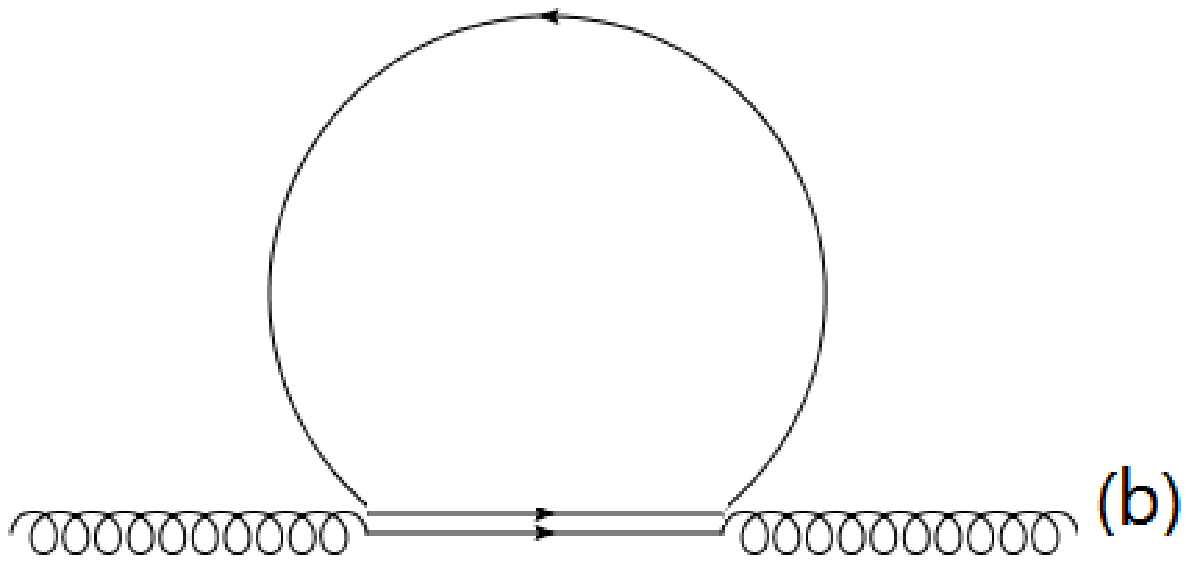}
\caption{Relevant one loop diagrams for the gluon rest masses. (a)
represents the ordinary polarization part. (b) represents the
momentum independent tadpole part which gives counterterm to
transverse gluon mass. Double line in (b) represents propagation in
Dirac sea mode \cite{Hong:1998tn, Hong:1999ru}.}\label{diagram2sc}
\end{figure}

While we expect the gluonic contribution to the symmetry energy in
the 2SC phase to be weak, it is worth while to estimate it. As we
have found in the previous section, contribution of HDL resummation
to the symmetry energy in the normal phase is determined dominantly
by the quark and gluon rest masses. Among them, the gluon
resummation with hard quark loop correction seems suitable for
physically relevant condition. In the 2SC phase, the gluonic
excitation will be soft or super-soft ($Q \sim T \ll g \mu$) and we
assume that the qualitative behavior of the symmetry energy in 2SC
phase also depends mainly on the quasi-quark loop correction in
small momentum limit. By using HDET \cite{Hong:1998tn, Hong:1999ru,
Casalbuoni:2001ha, Nardulli:2002ma}, the rest masses of gluon can be
simply calculated. In HDET, matter part of Lagrangian can be written
as follows \cite{Hong:1998tn, Hong:1999ru}:
\begin{align}
\mathcal{L}_q & = \sum_f \bar{\psi}_f(i \slash \hspace{-0.2cm}  D +
\mu_f \gamma_0) \psi_f\nonumber\\
& = \sum_{\vec{v}_f}\sum_{f}\left(\psi^\dagger_{f+}(i V\cdot D)
\psi_{f+} -\psi^\dagger_{f+} \frac{1}{2\mu_f+i\bar{V}\cdot D}
(\slash \hspace{-0.2cm}  D_\bot)^2 \psi_{f+} \right),
\end{align}
where $V^\mu=(1,\vec{v}_f)$ and $\bar{V}^\mu=(1,-\vec{v}_f)$ with
Fermi velocity $\vec{v}_f$ , $\slash \hspace{-0.2cm}
D_\bot=\gamma_\bot^\mu D_\mu$ with $\gamma_\bot^\mu = \gamma^\mu -
\gamma_\parallel^\mu$ and $\gamma_\parallel^\mu=(\gamma^0,\vec{v}_f
\vec{v}_f\cdot \vec{\gamma})$ and $\psi_{f+}$ represents positive
energy projection of quark field. The irrelevant quark mode
$\psi_{f-}$ is integrated out by using equation of motion:
$\psi_{f-} =-i\gamma^0/(2\mu_f+i\bar{V}\cdot D)\slash
\hspace{-0.2cm} D_\bot \psi_{f+}$.

\begin{table}
\addtolength{\tabcolsep}{+6pt}
\begin{tabular}{c |c c| c c| c c}
\hline\hline   &&$a,b=1,2,3$ & & $a,b=4,5,6,7$ & & $a,b=8$
\\[0.1cm]
\hline Paired (A=0)&& $\xi^{ab}_{AA}=\frac{1}{4}\delta^{ab}$ & &
$\xi^{ab}_{AA}=\frac{1}{8}\delta^{ab}$ &&
$\xi^{ab}_{AA}=\frac{1}{12}\delta^{ab}$ \\[0.1cm]
\hline Paired (A=1,2,3)&& $\xi^{ab}_{AA}=\frac{1}{4}\delta^{ab}$ & &
$\xi^{ab}_{AA}=\frac{1}{8}\delta^{ab}$ &&
$\xi^{ab}_{AA}=\frac{1}{12}\delta^{ab}$ \\[0.1cm]
\hline Unpaired (A=4,5)&&
$\xi^{ab}_{AA}=0$ & & $\xi^{ab}_{AA}=\frac{1}{4}\delta^{ab}$ &&
$\xi^{ab}_{AA}=\frac{1}{3}\delta^{ab}$ \\[0.1cm]
\hline\hline
\end{tabular}
\caption{List of $\xi^{ab}_{AA}$. Small letters a,b represent
adjoint color index of the external gluon. Capital letter A
represents the quasi-quark state in the loop of the tadpole diagram
(Fig.~\ref{diagram2sc}(b)).}\label{tbxi}
\end{table}

In 2SC phase, the quasi-quark states can be represented as follows
\cite{Casalbuoni:2001ha}:
\begin{align}
 \psi_{\pm,\alpha i} &= \sum_{A=0}^5
 \frac{(\tilde{\lambda}_A)_{\alpha i}}{\sqrt{2}}\psi_\pm^A,~~
 \chi^A =
  \left( {\begin{array}{c}
  \psi^A_{+}\\
  C\psi^{A*}_{-} \end{array}} \right),\nonumber\\
\tilde{\lambda}_{0} = \frac{1}{\sqrt{3}}
\lambda_{8}+\frac{2}{3}I;~\tilde{\lambda}_A &=
\lambda_A~(A=1,2,3);~\tilde{\lambda}_{4}=\frac{1}{\sqrt{2}}(\lambda_4-i\lambda_5);~\tilde{\lambda}_{5}=\frac{1}{\sqrt{2}}(\lambda_6-i\lambda_7),
\end{align}
where $A=0,1,2,3$ denotes gapped states, $A=4,5$ denotes ungapped
states and $\lambda_i$ is Gell-mann matrix. Here, `$\pm$' do not
represent the energy eigenstate. It represents direction of Fermi
velocity. Incorporating this representation and the gap interaction
term \eqref{2scil}, the Lagrangian can be written in Nambu-Gorkov
form \cite{Casalbuoni:2001ha, Nardulli:2002ma} as
\begin{align}
 \mathcal{L} =&-\frac{1}{4}F^a_{\mu\nu}F^{a\mu\nu}+ \sum_{\vec{v}_f}^{\textrm{half}} \sum_{A,B=0}^5\Bigg[ \chi^{A\dagger}
  \left( {\begin{array}{cc}
   i V\cdot \partial \delta_{AB} & \Delta_{AB} \\
   \Delta_{AB} &  i \bar{V}\cdot \partial \delta_{AB} \\
  \end{array} } \right)\chi^B +ig A^a_\mu  \chi^{A\dagger}
  \left( {\begin{array}{cc}
   i V^\mu \kappa_{AaB} & 0 \\
   0 &  -i \bar{V}^\mu \kappa_{AaB}^{*} \\
  \end{array} } \right)\chi^B\nonumber\\
   &+ g^2A^c_\mu A^d_\nu\chi^{A\dagger}
  \left( {\begin{array}{cc}
  \frac{1}{2\mu_f+i\bar{V}\cdot D}\xi^{cd}_{AB} & 0 \\
   0 &  \frac{1}{2\mu_f+iV\cdot D^{*}}\xi^{cd*}_{AB}\\
  \end{array} } \right)P^{\mu\nu}\chi^B\Bigg]+(L \rightarrow R),\label{lagrangian2sc}
\end{align}
where
$P^{\mu\nu}=g^{\mu\nu}-\frac{1}{2}(V^\mu\bar{V}^\nu+V^\nu\bar{V}^\mu)$
and the constants are defined as
\begin{align}
\kappa_{AaB}&
=\frac{1}{2}\textrm{Tr}[\tilde{\lambda}_A\tau_a\tilde{\lambda}_B],~~
\xi^{cd}_{AB}=\frac{1}{2}\textrm{Tr}[\tilde{\lambda}_A\tau_c \tau_d
\tilde{\lambda}_B],\nonumber\\
\Delta_{AB} &= \frac{\Delta}{2}\textrm{Tr}[\epsilon \sigma^T_A
\epsilon \sigma_B]~~(A,B=0,1,2,3,~\textrm{otherwise}~
\Delta_{AB}=0),\nonumber\\
&=\Delta_A
\delta_{AB}~~\textrm{with}~~\Delta_A=(-\Delta,\Delta,\Delta,\Delta,0,0),
\end{align}
where $\tau_a$ is the fundamental representation of SU(3) generator.

\begin{table}
\addtolength{\tabcolsep}{+3pt}
\begin{tabular}{c |c c | c c| c c}
\hline\hline   &&$a,b=1,2,3$ & & $a,b=4,5,6,7$ & & $a,b=8$\\[0.1cm]
\hline $\Pi_{0 0}^{ab}(0)$ && 0
&&$\frac{1}{2} m^2\delta^{ab} $ && $m^2\delta^{ab}$ \\[0.1cm]
\hline $-\Pi_{i j}^{ab}(0)$ && 0 &&
$\frac{1}{12}g^2\sum_{f}^{4,5}(\mu_f^2/\pi^2)\delta_{ij}\delta^{ab}$
&& $\frac{1}{9}[-m^2+g^2\sum_{f}^{4,5}(\mu_f^2/\pi^2)]\delta_{ij}\delta^{ab}$ \\[0.1cm]
\hline\hline
\end{tabular}
\caption{Debye and Meissner masses in the asymmetric 2SC matter.
Small letters a,b represent adjoint color index of the external
gluon. $m^2 = (g^2 \mu^2/\pi^2)$.}\label{DMmass}
\end{table}

For gluon self energy, the diagrams in Fig.~\ref{diagram2sc} are
relevant. The diagram Fig.~\ref{diagram2sc}(a) can be calculated
from ordinary current-current correlator
\begin{align}
\Pi_{\mu \nu}^{ab}(q)=g^2\int d^4 x e^{-iq\cdot x}
\sum_{\vec{v}_f}^{\textrm{half}}\left \langle  \Omega \vert T
[J_\mu^a(x,\vec{v}_f) J_\nu^b(0,\vec{v}_f)]\vert \Omega
\right\rangle,
\end{align} where $\vert \Omega\rangle$ denotes 2SC ground state and $J^\mu_a(x,\vec{v}_f)$ can be
obtained from the second term in the matter Lagrangian. The diagram
Fig.~\ref{diagram2sc}(b) directly comes from the last term in the
matter Lagrangian which has explicit iso-spin dependence. In
iso-spin symmetric condition, with counterterm
$\delta\Pi_{ij}^{ab}(0)=-\frac{1}{3}m^2\delta_{ij}\delta^{ab}$ which
should come from Fig.~\ref{diagram2sc}(b) type diagrams as it is
transverse component, the gluon rest masses can be matched with
Debye and Meissner masses calculated in Ref.~\cite{Rischke:2000qz}.
In iso-spin asymmetric condition, only thorough this counterterm,
iso-spin asymmetric part of the rest masses can be accounted.

The iso-spin asymmetric part in the whole counterterm
$\delta\Pi_{ij}^{ab}(0)$ can be determined by calculating
$\xi^{cd}_{AA}$. With $\xi^{ab}_{AA}$ in Table~\ref{tbxi}, the Debye
and Meissner masses in the asymmetric 2SC matter can be obtained.
The Debye and Meissner masses are listed in Table~\ref{DMmass}. From
Table~\ref{DMmass}, one can find that the rest masses of the gluon
in adjoint color $a=1,2,3$ vanishes. As this type of gluon carries
the unbroken SU(2) color charges which participate in the 2SC
condensation, the rest masses vanish trivially. The gluon in adjoint
color $a=4,5,6,7,8$ has rest masses. It can be noted that only the
Meissner mass is iso-spin weakly dependent. For the static quantity
such as thermodynamic potential, the contribution of the Debye mass
is most important. Consequently, in the 2SC phase, the contribution
of the Meissner mass to the symmetry energy will be minimal.
However, for dynamical quantities, the situation can be different.
The spatial transverse mode of super-soft gluon can acquire
nontrivial loop corrections  that can be as large as HDL
\cite{Arnold:1996dy, Huet:1996sh, Bodeker:1998hm, Bodeker:1999ud}.
This means that dynamical process which accompanies super-soft gluon
in 2SC phase may strongly depend on iso-spin asymmetry. Thus such a
process may critically influence observables that can discriminate
the existence of 2SC phase and identify the symmetry energy
properties of the cold dense matter.

\section{Discussion and Conclusions}

\begin{figure}
\includegraphics[height=7cm]{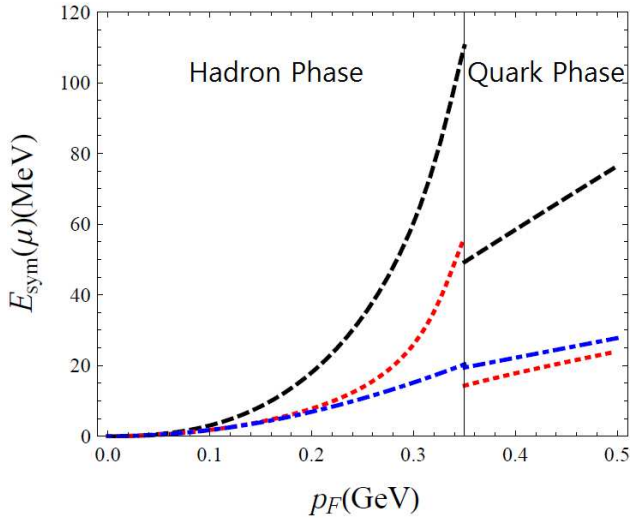}
\caption{(Color online) Symmetry energy as a function of the Fermi
momentum. Left side: symmetry energy in the hadron phase calculated
by using QCD sum rule \cite{Jeong:2012pa}. Blue dot-dashed line
represents the kinetic part of symmetry energy of the ideal nucleon
gas. Red dotted line represents the kinetic part of interacting
matter. Black dashed line represents the total symmetry energy
(kinetic + potential). Right side: quark matter symmetry energy in
the normal and 2SC phase. Blue dot-dashed line represents symmetry
energy of ideal quark gas. Red dotted line represents symmetry
energy in the normal phase with HDL resummation. Black dashed line
represents symmetry energy in the 2SC phase. The phase boundary has
been set artificially.}\label{qh}
\end{figure}

In this work, we calculated the thermodynamic potential and the
symmetry energy at the cold dense limit. For the normal phase at low
but non-negligible temperature, we first calculated the free energy
by using HDL resummation, which corresponds to calculating quark and
gluon ring diagrams with HDL self energy. From a theoretical point
of view, the gluon resummation seems to be more suitable for the
physical low energy excitations in the cold dense matter. For the
quark resummation, we optionally considered the effects of the
resummation. All the thermodynamic properties,  calculated from the
thermodynamic potential, strongly depends on the quark and gluon
rest mass. In the gluon self energy as well as the free energy
itself, the largest contribution comes from the Debye mass of the
longitudinal mode, which comes mainly from the quark-hole loop in
HDL self energy, whereas the contribution from the transverse mode,
which comes from quark anti-quark loop, is comparatively
unimportant. The symmetry energy also comes mainly from the iso-spin
dependence of the Debye mass. HDL resummed gluonic interaction gives
reducing effect to the symmetry energy.

For the 2SC phase, which can appear when temperature becomes
extremely low, the free energy containing non-perturbative
condensation should be considered. Following previous studies
\cite{Schafer:1999fe, Miransky:1999tr}, the free energy can be
written in a simplified form as given in Eq.~\eqref{g2sc}, from
which one can extract the symmetry energy in the 2SC phase by
reducing the formula to have the iso-spin asymmetric factor
$I_{\tilde{B}}$. The symmetry energy becomes almost 3 times that in
the normal phase, because  2-color and 2-flavor states are locked in
their common Fermi sea,  reducing the number of available quarks
that can contribute to the asymmetry. Although the symmetry energy
becomes larger than the typical value in the normal quark phase, it
still remains smaller compared to the expected value in the hadronic
phase extrapolated to the phase boundary (Fig.~\ref{qh}).  We also
estimated the gluonic contribution to the symmetry energy in the 2SC
phase. By using HDET \cite{Hong:1998tn, Hong:1999ru,
Casalbuoni:2001ha, Nardulli:2002ma}, we found that only the Meissner
mass in the 2SC phase has a small iso-spin dependence
(Table.~\ref{DMmass}).  In another words, as far as the gluons are
concerned,  the Debye mass is important for the static quantities,
and  Meissner mass contributes to the symmetry energy, but with an
even smaller reducing effect than in the normal phase.   Hence,  one
can just consider the effect of BCS condensation only.

With the results on the cold dense matter symmetry energy in hand,
one can imagine a phenomenological scenario in which one can create
a cold dense matter through a heavy ion collision. In Fig.~\ref{qh},
one finds that the nuclear symmetry energy in the hadronic phase
rises stiffly. The nuclear symmetry energy was calculated by using
QCD sum rules, based mainly on linear density approximation
\cite{Jeong:2012pa}. The OPE terms and formula for plotting density
the behavior can be found in Ref.~\cite{Jeong:2012pa}. Similar stiff
density behavior of the symmetry energy was reported before in
Ref.~\cite{Baran:2004ih}, including iso-spin dependent interaction
channel (NL$\rho\delta$). If the symmetry energy increases stiffly
at high density region, the neutron sea will become unstable. Then,
it may lead to $nn\rightarrow p\Delta^{-}$ type scattering. As the
unstable neutrons are scattered out, the iso-spin asymmetry of
remaining nuclear matter will be reduced. But, as proposed in
Ref.~\cite{Di Toro:2006pq, DiToro:2009ig}, if the mixed phase with
quark and hadron appears, the iso-spin asymmetry can remain  high
due to the small symmetry energy of the quark phase: iso-spin
distillation, by forming deconfined quark drop in the neutron. This
distillation effect will cause enhanced $\pi^{-}$ yields compared to
the case where there is only hadronic phase. Our calculation shows
enhanced quark matter symmetry energy when 2SC phase appears. So the
distillation effect will be reduced and the ratio $\pi^{-}/\pi^{+}$
will be smaller than the case where only the normal phase exists
\cite{Di Toro:2006pq, DiToro:2009ig}. Our consideration is in
agreement with Ref.~\cite{Pagliara:2010ii}. If the heavy ion
collision experiments can be arranged to produce different
temperature of dense matter, these two trend can be identified: one
goes to  the normal quark matter with low temperature and the other
to the 2SC quark matter with extremely low temperature. The two
experiment may give enhanced ratio $\pi^{-}/\pi^{+}$ compared to the
hadronic phase, but with a quite different value. Such experiment
may be available at FAIR \cite{Stocker:2015cva}.

Although we have neglected strangeness degree of freedom for
simplicity, considering its effects may give us a fruitful insight.
Even if they are not the main excitation in the matter, one can
expect high multiplicity of the hadron excitation containing
strangeness. Measuring kaon multiplicity may be good probe for the
cold dense matter as it has higher threshold energy and do not
strongly interact with surrounding nuclear matter
\cite{Aichelin:1986ss, Fuchs:2005zg}. Also the iso-spin dependence
of hyperon mass and its interaction scattering cross section at high
density regime can be an interesting issue. Describing these physics
using  explicit QCD quantum number may provide new understandings
for high density iso-spin asymmetric matter. Related study is now in
progress.

\begin{acknowledgements}
This work was supported by Korea national research foundation under
Grants No. KRF-2011-0030621 and No. KRF-2011-0020333.
\end{acknowledgements}

\end{document}